\documentclass{article} 
 
\title{The Goldberg-Sachs theorem in linearized  gravity}

\author{Sergio Dain\thanks{Fellowship holder  of C.O.N.I.C.O.R} 
\thanks{E-mail: dain@aei-potsdam.mpg.de} 
\thanks{Present address: Max-Planck-Institut f\"ur Gravitationsphysik,
  Am M\"uhlenberg 1, D-14476 Golm, Germany}\, and Osvaldo
M. Moreschi\thanks{Member of CONICET}\\
FaMAF, Ciudad Universitaria, Universidad Nacional de C\'ordoba, \\
(5000) C\'ordoba, 
Argentina}
 
\usepackage{amsmath,amssymb,amsfonts} 
 
\newtheorem{theorem}{Theorem}

\begin{document}

\maketitle 
 
\begin{abstract} 
The Goldberg-Sachs theorem has been very useful in  
constructing   algebraically special exact solutions of Einstein vacuum equation. 
Most of the physical meaningful vacuum exact solutions are  
algebraically special.  
We show that the  Goldberg-Sachs theorem is not true  
in linearized  gravity.  
 
This is a remarkable  result, which gives light on the understanding 
of the physical meaning of the linearized solutions. 
 
\vspace{0.5cm} 
\noindent PACS numbers: 04.20-q, 04.25-g 
  
\end{abstract}

Solutions of the linearized Einstein vacuum equation are usually  
considered as approximations of solutions of the full vacuum equation. They are useful tools for describing physical systems.  
 
It is important to understand the relation between the full vacuum  
equation and the linearized one; in particular it is interesting to  
 know which  
properties are common, or not, to both sets of solutions. 
An example of a common property  is  the Birkhoff's theorem; which  can be enunciated in the following  
 way: vacuum spherically symmetric solutions are static. This statement 
 remains true in linearized gravity (i.e., if we replace the vacuum equation by the linear one); as one can check by reconstructing  
the linear version of the  standard proofs  that appear in  
the literature.

In this work  we present an example of a property, the so called  
Goldberg-Sachs theorem\cite{Goldberg62},  which is not  
 common to both set of solution.  
 
The Goldberg-Sachs theorem for the Einstein vacuum equation relates   
algebraic  properties of the Weyl tensor with the existence of a  
null, geodesic, shear-free congruence in the space-time. In the search of vacuum solutions, the existence of such a congruence  
leads to considerable simplification in the calculation. The  
 Schwarzschild, Kerr and Robinson-Trautman space-times, which are probably the most useful metrics in the study of compact objects, are all algebraically special.  
However our present result shows that  if one has a 
linearized solution admitting shear free null geodesic  congruence, then in general it  will  not be  algebraically special. The proof of this  consists  in presenting  explicit counterexamples.

The statement of  Goldberg-Sachs theorem is\cite{Goldberg62} (see also \cite{Kramer80} \cite{Penrose86})  
\begin{theorem}[Goldberg-Sachs] 
A vacuum metric admits a  shear free null geodesic congruence $l^a$ if and only if  $l^a$ is a degenerate eigendirection of the Weyl tensor 
\begin{equation} \label{Alsp}
C_{abc[d}l_{e]} l^b l^c = 0.
\end{equation}
\end{theorem}
The condition (\ref{Alsp}) is what characterize algebraically special space-times. 
 
Let us study  this statement in the context of linearized  gravity. 
Consider  a one-parameter family of metrics $g_{ab}(\gamma)$ with the corresponding metric, torsion free, connection $\nabla_a$, Weyl tensor $C_{abcd}$ and Ricci tensor $R_{ab}$. We define the first order Weyl and Ricci tensor by 
\[
C^{(1)}_{abcd}=\frac{d}{d\gamma}C_{abcd}|_{\gamma=0}, \quad R^{(1)}_{ab}=\frac{d}{d\gamma}R_{ab}|_{\gamma=0}. 
\]  
In the same way we can define the `first order' shear of a vector field, and so on. 
  
We assume  that  
\begin{equation} \label{fflat} 
g_{ab}(\gamma=0)= \eta_{ab}, 
\end{equation} 
where $\eta_{ab}$  is the flat metric. We say that $g_{ab}(\gamma)$ satisfies the linear vacuum equation if 
\[ 
R^{(1)}_{ab}=0. 
\] 
  
Let us assume that $g_{ab}(\gamma)$ satisfies the linear vacuum equation, and also that  $g_{ab}(\gamma)$ admits a first order geodesic shear free null  vector field $l^a$. We ask the following question:   Is it true that   $l^a$ is a degenerate eigendirection of the first order Weyl tensor ($C^{(1)}_{abc[d}l_{e]} l^b l^c = 0$)? Note that this question is a linearized version of the first  implication of the Goldberg-Sachs theorem. 
 
We  prove  that the answer is negative  by giving an explicit one-parameter family $g_{ab}(\gamma)$ and a vector field $l^a$,  with the following properties (we define $l_a\equiv g_{ab}l^b$): 
 
\begin{itemize} 
  
\item[(i)] The vector field $l^a$ is null, geodesic and shear 
free with respect to $g_{ab}(\gamma)$, i.e., it  satisfies the next equations 
\begin{align*} 
l^a l^b g_{ab}(\gamma)&=0,\\ 
l^a \nabla_a l^b&=\phi l^b,\\   
\nabla_{(a} l_{b)}&=J_{(a}l_{b)}+ \theta g_{ab}, 
\end{align*}  
for some scalar fields $\phi$, $\theta$ and some vector field $J_a$. The last equation means that the  shear is zero. 
 
\item[(ii)] The family $g_{ab}(\gamma)$ satisfies the linear field equation 
\[ 
R^{(1)}_{ab}=0. 
\] 
 
\item[(iii)] The vector field $l^a$ is not a degenerate eigendirection of the linear Weyl tensor 
\[ 
C^{(1)}_{abc[d}l_{e]} l^b l^c \neq 0. 
\] 
 
\end{itemize} 
 
We would like to emphasize here that our example below satisfies  (i) 
exactly; i.e., to all orders in $\gamma$. 
 Of course in order to construct a counter-example it is enough to satisfy this  condition only up to  first order in $\gamma$.

A one-parameter family $g_{ab}(\gamma)$  which satisfies  condition  
(i), can be constructed in the following way. 
 
Let $l^a$ be a vector field that is null, geodesic, and shear-free  
with 
respect to the flat connection $\partial_a=\nabla_a(\gamma=0)$: 
\begin{align}  
l^al^b\eta_{ab}&=0,\label{flatconditions1}\\ 
l^b \partial_b l^a&=\phi l^a, \label{flatconditions2} \\ 
 \partial_{(a} l^0_{b)}&=J_{(a}l^0_{b)}+\theta \eta_{ab},\label{flatconditions3} 
\end{align} 
for some scalars fields $\phi$ and $\theta$, some vector  
field $J_a$, and where we have defined $l^0_a=\eta_{ab}l^b$ in order to distinguish it  from $l_a=g_{ab}l^b$. 
In other words, $l^a$ satisfies  condition (i) for $\gamma=0$.  
 
Consider 
the following one-parameter family  
\begin{equation}\label{Gform} 
g_{ab}(\gamma)=\eta_{ab}+\gamma (\mu\eta_{ab}+l^0_{(a}v_{b)}), 
\end{equation} 
where $v_b$ is a  
smooth vector field, and $\mu$ a smooth scalar field.  
 
We now claim:  For all $v_a$ and $\mu$, the metric  $g_{ab}(\gamma)$ satisfies condition (i), i.e.,  it preserves the null, geodetic,  
shear-free character of $l^a$.

First note that preservation of the null character 
\begin{equation}\label{gnull} 
l^al^bg_{ab} = 0 
\end{equation} 
is immediate.  
 
Now we compute $l^a \nabla_a l^b$. We  calculate the Lie derivate  $\pounds_l$ of $l^0_a$ 
\[ 
\pounds_l(l^0_a)=l^b \partial_b (l^0_a)+ l^0_b \partial_a l^b, 
\] 
using  equations   (\ref{flatconditions1}) and (\ref{flatconditions2}) we obtain 
\begin{equation} \label{Liel0} 
\pounds_l(l^0_a)=\phi l^0_a . 
\end{equation} 
The Lie derivative can also be written in terms of the connection $\nabla_a$: 
\begin{equation} \label{Liegl0} 
\pounds_l(l^0_a)=l^b \nabla_b  l^0_a+ l^0_b \nabla_a l^b=\phi l^0_a. 
\end{equation} 
We use (\ref{Gform}) to replace $l^0_a$ by  
\begin{equation} \label{l0} 
l^0_a=l_a (1+\gamma(\mu+l^cv_c))^{-1} 
\end{equation} 
in  equation (\ref{Liegl0}) to  obtain 
\begin{equation} \label{ggeodesic} 
l^b \nabla_b l^a= l^a\left( \frac{ l^c\partial_c \gamma (\mu+l^ev_e)}{1+\gamma (\mu+l^ev_e)}+\phi \right) ; 
\end{equation} 
where we have used equation (\ref{gnull}). Then $l^a$ is geodesic with respect to  
$g_{ab}$. 
 
 It remains to be shown  that $l^a$ is shear-free with respect to $g_{ab}$. We have to prove that 
 $\nabla_{(a} l_{b)}$   is equal to some vector symmetrized with $l_a$ plus some  
multiple of $g_{ab}$.  
In order to  prove this, let us  note that  
\begin{equation}  
\begin{split} 
\label{Lieshear} 
2 \nabla_{(a}l_{b)} =&\pounds_l g_{ab}\\ 
=&\gamma\eta_{ab} \pounds_l \mu+(\pounds_l v_{(a})l^0_{b)}\\ 
&+(1+\gamma \mu)\pounds_l (\eta_{ab})+( \pounds_l l^0_{(a}) v_{b)}. 
\end{split} 
\end{equation} 
 
We will prove that each term in  the right-hand side of equation (\ref{Lieshear}) has the desired form.  
 
For the first and second term  one  only has to  use the definition  
(\ref{Gform}) and the relation (\ref{l0}). 
For the third term one  uses that     $\pounds_l (\eta_{ab})=2\partial_{(a}l_{b)}$ and   equation  (\ref{flatconditions3}) (the flat shear-free condition). Finally, for the last term  one  uses  
equation (\ref{Liel0}).

Consider now a  metric of the following  form 
\begin{equation} \label{KS} 
g^{KS}_{ab}(\gamma)=\eta_{ab}+\gamma fl_al_b, 
\end{equation} 
where  $f$ is some scalar field.  
 We assume also that $g^{KS}_{ab}(\gamma)$ satisfies the linear 
vacuum equation (condition (ii) ).  
This class  of metrics are said to have the  Kerr-Schild form\cite{Kerr65}\cite{Debney69}. Examples of these  metrics are the Schwarzschild and Kerr metrics. 
 
Let $k^a$ be an arbitrary Killing vector of  $\eta_{ab}$ (i.e. it satisfies  $\pounds_k \eta_{ab}=0$). Take the metric given by  
\begin{equation} \label{hatKS} 
g_{ab}(\gamma)=\eta_{ab}+\pounds_k g^{KS}_{ab}. 
\end{equation} 
Since by hypothesis $g^{KS}_{ab}$ satisfies the linear vacuum equation, it follows that also   
$g_{ab}$ satisfies it (condition (ii)); and also we can see that  condition (i) is fulfilled; since one can easily show that it has 
the form (\ref{Gform}).  
 
The linear Weyl tensor corresponding to $g_{ab}$ is 
\begin{equation} \label{Weyl} 
C^{(1)}_{abcd}=\pounds_k C^{KS(1)}_{abcd},  
\end{equation} 
where  
$C^{KS(1)}_{abcd}$ is  the  linear Weyl tensor of $g^{KS}_{ab}$. 
 
One can prove that the vector $l^a$ is a degenerate eigendirection of  $C^{KS(1)}_{abcd}$. But, if we choose $k^a$  such that $\pounds_k l^a$ is not proportional to $l^a$, then $l^a$  
 will be a principal null direction of $C^{(1)}_{abcd}$ but not a degenerate one; as one can  see from equation (\ref{Weyl}). An explicit  
example of this situation is given when  $g^{KS}_{ab}$  is  the Schwarzschild metric and $k^a$ is any space translation. 
 
Then the metric $g_{ab}$ satisfies also condition (iii), and this completes the proof.

The given counterexamples are  by no means  the only possible ones.  
In a separate  work we discuss another  class of linearized solutions, 
which include also different counterexamples to the Goldberg-Sachs theorem in linearized gravity\cite{Dain00}.

Given an exact vacuum solution which depends smoothly on some 
parameter $\gamma$ whose vanishing yields Minkowski space, and which 
is algebraically special for all values of the parameter, then the  
corresponding linearized 
 solution will also be algebraically special. The solutions we indicate in 
  this work belong to a set of linearized solutions that can not be reached  
 by this means. They are linearizations of some vacuum solutions which 
 are not algebraically special.  These  vacuum solutions have a vector field 
 $l^a$ which is null, shear free and geodesic only up to the first 
 order in  $\gamma$,  but they  are not algebraically special not even 
 up to the the first order in   $\gamma$.

We have here shown  that the Goldberg-Sachs theorem is  
one of those properties of the Einstein vacuum equation      
 with no analog in the linear theory. As we have mentioned in the introduction, most useful metric in the study of compact object are algebraically special;  
we hope that our result will  contribute to understand the   
relevance  of the algebraic special condition  
 in the physically  
meaningful solutions of  Einstein vacuum equation.

\section*{Acknowledgments} 
We are deeply indebted  to R. Geroch for  very illuminating  
discussion related to this work, in particular we thank  him for  
indicating to us how to construct the counterexample using  
the Kerr-Schild metrics. 
 
We acknowledge support from  Fundaci\'on Antorchas, SeCyT-UNC, CONICET and 
FONCYT BID 802/OC-AR PICT: 00223.

\end{document}